\documentclass[runningheads,fleqn]{cl2emult}

\usepackage{makeidx}  % allows index generation
\usepackage{graphicx} % standard LaTeX graphics tool
\usepackage{subeqnar} % subnumbers individual equations
\usepackage{multicol} % used for the two-column index
\usepackage{cropmark} % cropmarks for pages without
\usepackage{phys}     % flushleft layout of math and captions
\usepackage{epsfig}
\makeindex
\def\PRL{Phys. Rev. Lett.\ }

\begin{document}

\title*{Macroscopic Finite Size Effects in Relaxational Processes} 
%%\titlerunning{Macroscopic Finite Size Effects in Relaxational Processes} 

\author {Shlomo Havlin\inst{1,2}
\and Armin Bunde\inst{1,2}
\and Joseph Klafter\inst{3}}
%%\authorrunning{Shlomo Havlin et al.}

\institute{Institut f\"ur Theoretische Physik III,
Justus-Liebig-Universit\"at Giessen, D-35392 Giessen, Germany
\and Minerva Center and Department of Physics,
Bar-Ilan University, Ramat Gan, Israel
\and School of Chemistry, Tel Aviv University, Tel Aviv, Israel}

\maketitle

\begin{abstract}
We present results on dynamical processes that exhibit a stretched
exponential relaxation.  When
the relaxation is a result of     two competing
exponential processes, the size of the system, although macroscopic,
 play a dominant role.  There exist a crossover time $t_\times$ that
depends
{\it logarithmically\/} on the size of the system, above which, the
relaxation changes from a stretched exponential to a simple exponential
decay.  The decay rate also depends {\it logarithmically\/}
on the size of the system.  The results are relevant to
large-scale Monte-Carlo simulations and should be amenable to
experiments  in low-dimensional macroscopic systems and mesoscopic
systems.
\end{abstract}

\section*{}
Many relaxational
processes in macroscopic systems are characterized by a
relaxation function $Q(t)$ that exhibits a stretched exponential 
behavior,
\begin{equation}
Q(t)\sim Q(0)\exp[-(t/\tau)^{\beta}],\label{eq1}
\end{equation}
where  $0<\beta<1$.  Examples include viscoelastic relaxation
\cite{bib01}, dielectric relaxation \cite{bib02}, glassy relaxations
\cite{bib03,bib04,bib05},
relaxation in polymers \cite{bib06,bib07} and long-time decay in trapping
processes \cite{bib08}.  Many more examples
\cite{bib09,bib10,bib11,bib12,bib13,bib14,bib15} suggest that (\ref{eq1})
is common to a very wide range of phenomena and macroscopic
materials.

The origin of the stretched exponential is not always clear. 
In many cases it is assumed to be the result of a competition
between two exponential processes.  In some cases, e.g., trapping
processes at long times, this assumption is well established,
while in others, such as relaxation in glassy materials, this
assumption has been controversially discussed \cite{bib16,bib17} and
alternative models have been also suggested
\cite{bib10,bib18,bib19,bib20}.

We have recently investigated the occurence of stretched exponential
behavior  in  finite systems, in cases where the relaxation arises due to two
competing  exponential processes 
\cite{bib21}                                                     

We have found that:                                 
(a) the size of the system, although macroscopic, plays a dominant role in 
the relaxation time        
pattern, leading to an exponential decay sufficiently at long times;
(b) the crossover time, $t_\times$, to the exponential depends
logarithmically on the system size;
(c) the  rate of the exponential decay also depends logarithmically on the     
system size, and (d) in the special examples of the trapping and the                  
hierarchically constrained dynamics models the exponential relaxation may      
enter before the stretched exponential is reached. 
These results are of
relevance to experiments in confined systems, mesoscopic systems and to
Monte-Carlo simulations.
Our theoretical predictions on the finite size effects can serve as an
experimental test for identifying the origin of the mechanism
leading to stretched exponential decay.

  We assume that   the relaxation function of the whole
system can be represented by an integration over all possible
states $n$, namely,
\begin{equation}
Q(t)=\int_0^{\infty}\Phi(n) Q(n,t)dn.\label{eq2}
\end{equation}
Here, $\Phi(n)$ is the probability that state $n$ is occupied
 and $Q(n,t)$ is the dynamic relaxation of the
 $n$-th state.

Usually, in the case of
a stretched exponential behavior,
$\Phi(n)$ is assumed to behave as 
$\Phi(n)\sim\exp(-an^{\alpha})$, while $Q(n,t)$ decays 
exponentially with time as $Q(n,t)\sim\exp(-bt/n^{\gamma})$.
A number of dynamical models that yield a stretched
exponential decay can be formulated in terms of Eq.(\ref{eq2}).
These include the long-time behavior in the trapping
problem \cite{bib08}, the target problem \cite{bib20}, direct energy
transfer \cite{bib20}, trapping of nonidentical interacting
particles \cite{bib23},
hierarchically constrained dynamics \cite{bib16},
models for relaxation in microenulsions and molecular glasses \cite{bib24}
and others.

We now concentrate on  three examples: 
(i) A particle diffusing in a $d$-dimensional system with
randomly distributed static traps, where we are interested in the
survival probability $Q(t)$ of a particle.
Here the state $n$ represents a particle in a trap-free
region of linear size $n$; $\Phi(n)$ is the probability for
the occurance of a size $n$ trap-free region, and $Q(n,t)$
is the survival probability of the particle in this region \cite{bib08}.
The exponent $\alpha$ is the dimension $d$ of the system,
and $\gamma=2$ due to the diffusional motion.  
(ii) A linear system (chain) along which two types of particles ($A$ and $B$)
are diffusing and interacting
via hard core interaction. However, only type $A$ can be
trapped by  static traps which are randomly distributed along the
chain.
Here, $Q(t)$ is the survival probability of particles of type $A$,
$\Phi(n)$ is the probability that a free trap region of size $n$ occurs,
and $Q(n,t)$ is the survival probability of a type $A$ particle to survive
in this region. The exponent $\alpha$ is the dimension of the system $\alpha=1$ and
$\gamma=4$ is due to diffusion in the presence of hardcore iterations
\cite{bib25}.
(iii) 
Hierarchically constrained dynamics, a model that has been
proposed to account for glassy relaxation \cite{bib16}. 
This
model assumes that the relaxation of level $n$ populated
by spins, occurs in stages, and
the constraint imposed by a faster degree of freedom must
relax before a slower degree of freedom can relax.  This
implies that the time scale of relaxation in one level is
subordinated to the relaxation below.  A possible realization
considered in \cite{bib16} and here is a system with a discrete
series of levels where the relaxation time of level $n$ 
 is
$\tau_n\sim n^\gamma$ (corresponding to the exponential
form of $Q(n,t)$ in (\ref{eq2})), and the weight factor of level $n$,
is $\Phi(n)\sim e^{-an}$ \cite{bib12}, corresponding to $\alpha=1$.
   The first
exponential in (\ref{eq2}) is accordingly the probability to occupy level
$n$ and the second exponential represents the decay of that level.  

The evaluation of the long time behavior of the integral in
(\ref{eq2}) is performed using the method of steepest descent.  The main
contribution to the integral arises from the maximum of the
integrand in (\ref{eq2}), which is obtained from the minimum of the
function, $-an^{\alpha}-bt/n^{\gamma}$, appearing in the
exponent.  This yields that the main contribution to (2)
comes from
\begin{equation}
n^{*}\cong (\gamma b t/\alpha a)^{1/(\alpha+\gamma)},\label{eq3}
\end{equation}
leading to \ref{eq1} with  $\beta=\alpha/({\alpha+\gamma})<1$,
and   $\tau=(\alpha/b\gamma) a^{-\gamma/\alpha}(\gamma/(\gamma+\alpha))^
{1+\gamma/\alpha}$.

However, as shown below, these arguments are valid only in the
thermodynamic limit where the system size is infinite.   For a
finite number $N$ of traps (in the trapping system) or a
{\it finite\/} system with a finite number $N$ of spins (in the
hierarchical constraint system) the relaxation function depends explicitly
on $N$.  Since our discussion is quite general for  systems described
by (\ref{eq2}), in what follows we refer below to traps and spins in the
above examples as elements.

 For  a single finite  
system consisting of $N$ elements, the relaxation function $Q(t)$
represents   an 
average quantity over the $N$ elements,
\begin{equation}
Q(t)={1\over N}
\sum_{\{n\}}m(n)Q(n,t),\label{eq4}
\end{equation}
  where  the sum is over all possible states $n$ and $m(n)$
is the number of elements at state $n$, with $\sum_{\{n\}}m(n)=N$.
  Since the sum in (\ref{eq4}) is over exponential functions, the value
of $Q(t)$ will fluctuate for different sets of $N$.
There will be a distribution of $Q(t)$, and we are
interested in the typical  $Q(t)$, which is around the peak of this
distribution.  

In the thermodynamic limit
$N\to\infty$, all states $n$ are occupied,
$m(n)/N$ can be identified with $\Phi(n)$ and (\ref{eq2}) follows.
For $N$ finite, in contrast, there exists a
characteristic  "maximum" state 
$n=n_{\rm max}(N)$, and this $n_{\max}$
should replace the upper limit ($\infty$) in (\ref{eq2}),
\begin{equation}
Q(t)=\int_0^{n_{\max}} \Phi (n)Q(n,t)\,dn. \label{eq5}
\end{equation}

  To estimate how $n_{\max}$ depends on $N$,
we note that  the typical number of states $n$
  in a sample of $N$
elements    is $Z(n)\cong N \Phi(n)\cong N
 \exp(-an^{\alpha})$. States with $Z(n)\ll 1$
 will not occur in a typical system of $N$ elements, and this yields  
\begin{equation}
n_{\max}\cong \left({\ln N\over a}\right)^{1/\alpha}.\label{eq6}
\end{equation}

If $n^*\ll n_{\max}$, the upper limit in (\ref{eq2}) can be approximated
by infinity and  thus leads to (\ref{eq1}).  However,
if $n^*\gg n_{\max}$ the main contribution to (\ref{eq5}) will
not be from the maximum of the integrand, which is 
outside the range of integration, but from $n_{\max}$.
Thus, for  $n^*\gg n_{\max}$ we expect
\begin{equation}
Q(t)\cong Q(0)e^{-bt/n_{\max}^{\gamma}}\label{eq7}
\end{equation}
where the time constant of the relaxation, $n_{\rm max}^\gamma$,
scales as $(\ln N)^{\gamma/\alpha}$.
The crossover time from a stretched exponential  (\ref{eq1})
to an exponential (\ref{eq7}) can be estimated from the condition
 $n^*=n_{\max}$, from which follows
\begin{equation}
t_{\times}\cong{\alpha a\over \gamma b}
\left({\ln N\over a}\right)^{1+\gamma/\alpha}.\label{eq8}
\end{equation}
The striking point in (\ref{eq8}) is the logarithmic dependence
on $N$, which puts $t_{\times}$ in the range of observable
time scales measurable in mesoscopic and even macroscopic systems.
Indeed, the corresponding   relaxation   value $Q(t_\times)$ scales as
\begin{equation}
Q(t_\times)\sim N^{-\alpha/\gamma},\label{eq9}
\end{equation}
independent of the microscopic parameters $a$ and $b$. For the above
three cases we find:
(i) In the case of the trapping relaxation mechanism where $\alpha=d$
and $\gamma=2$ we obtain,
\begin{equation}
Q(t_\times)/Q(0)\sim N^{-d/2}.\label{eq10}
\end{equation}
(ii) In the non identical case $\alpha=1$ and $\gamma=1$ and thus
\begin{equation}
Q(t_x)/Q(0)\sim N^{-1/4}.\label{eq11}
\end{equation}
(iii) In the hierarchical constraint dynamics
\begin{equation}
Q(t_\times)/Q(0)\sim N^{-1/\gamma}.\label{eq12}
\end{equation}
  It is known [8e,23] %%\cite{8e,23}
that in both examples, for an infinite system,
the stretched exponential behavior of (\ref{eq1}) sets in only at very long
times.. Thus we expect that in the finite system, the crossover will mask
the stretched-exponential pattern. 

To test our analytical approach, we performed new Monte Carlo
simulations on two cases (i) and (iii), the trapping model (case (i)) and 
the hierarchical constraint model (case (iii)).
In the trapping model, we consider one and two dimensional systems with
a fixed concentration $c=0.5$ of randomly distributed traps,
and vary the size $N/c$ of the system. We calculated numerically the
survival probability $Q(t)$ of a particle as a function of $t$ and
$N$. 
In the hierarchical model we have chosen $\tau_n\sim n$ i.e., $\gamma=1$.
We calculated the relaxation function 
for system sizes varying from $N=10^2$ to $N=10^5$.
  
As mentioned earlier, the relaxation function fluctuates for
different sets of $N$. For obtaining the typical behavior
of $Q(t)$, we have considered therefore the "typical"
 average $Q(t)_{\rm typ}\equiv\exp(\langle\ln Q(t)\rangle$,
where the brackets denote an average over many sets of $N$ elements.
Note that an arithmetic average over $M$ sets of $N$ elements can not be employed
here, since it leads to a result identical for a larger system with
$M\times N$ elements (see \cite{bib04}). For a discussion of typical
averages see \cite{bib26}.
For simplicity, we shall drop the index "typ" in the following.
 
\begin{figure}
\epsfig{file=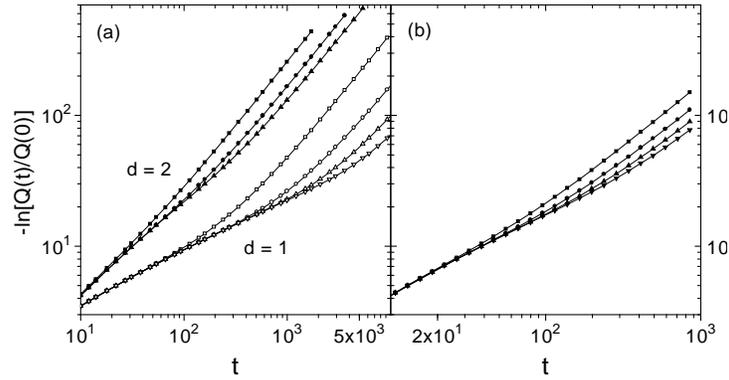,width=6cm,clip=,bbllx=120,bblly=140,bburx=450,bbury=740,angle=-90}
\caption{Plot of $-{\ln[Q(t)/Q(0)]}$ as a function of $t$ in
a double logarithmic presentation for ({\bf a}) the trapping model
in $d=1$ and $d=2$, and ({\bf b}) the hierarchical constraint
model, for several system sizes.
For the trapping model, the system sizes are $N=2\cdot10^3$ (open square),
$2\cdot10^5$ (open circle), $2\cdot10^7$ (open up triangle), $2\cdot10^9$ (open
down triangle) in $d=1$, and $N=9\cdot10^2$ (full square), $9\cdot10^4$ (full
circle), $9\cdot10^6$ (full up triangle) in $d=2$.
For the hierarchical model, the system sizes are $N=10^2$ (full square), $10^3$
(full circle), $10^4$ (full up triangle), $10^5$ (full down triangle).}
\label{fig01}
\label{fig01a}
\label{fig01b}
\end{figure}

Figure 1 shows $-{\ln[Q(t)/Q(0)]}$ as a function of $t$ in
a double logarithmic plot for (i) the trapping model
in $d=1$ and $d=2$, and (iii) the hierarchical constraint
model, both for several system sizes. In all cases, a crossover 
from an exponent $\beta < 1$ (at small $t$) towards $\beta=1$ (at large $t$)
can be easily recognized.
The crossover time $t_\times$ shifts towards larger values
when $N$ increases.

\begin{figure}
\epsfig{file=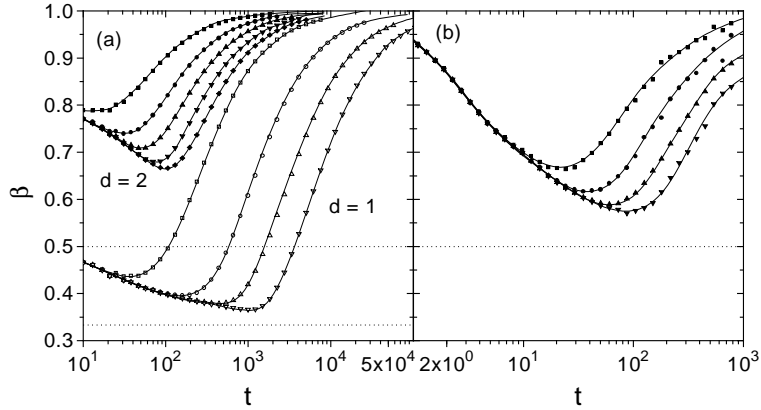,width=6cm,clip=,bbllx=140,bblly=210,bburx=450,bbury=760,angle=-90}
\caption{Plot of the local exponents $\beta$ calculated from the successive
slopes of the corresponding curves in (\ref{fig01}), ({\bf a}) 
 for the trapping model and ({\bf b}) for the
 hierarchical model. The horizontal dashed lines represent the 
 corresponding asymtotic
($N\to \infty, \quad t\to \infty$) values of $\beta$.}
\label{fig02}
\label{fig02a}
\label{fig02b}
\end{figure}
 
To study the crossover behavior in a more quantitative manner,
we have plotted in Fig.~\ref{fig02} the local exponents $\beta$ obtained from
the local slopes of Fig.~\ref{fig01}, as a function  of $t$. In both systems, 
for a fixed system size $N$,  $\beta$ first decreases with $t$, reaches
a minimum value at a certain time that can be identified with $t_\times$,
and then increases monotonically with time towards $\beta=1$. The figure shows
that the minimum value of $\beta$ has not yet reached its asymptotic
value predicted for infinite systems, i.e., $\beta=1/3$ ($d=1$)
 and $\beta=1/2$ ($d=2$) for the trapping system and $\beta=1/2$ for the 
 hierarchical system.

\begin{figure}
\epsfig{file=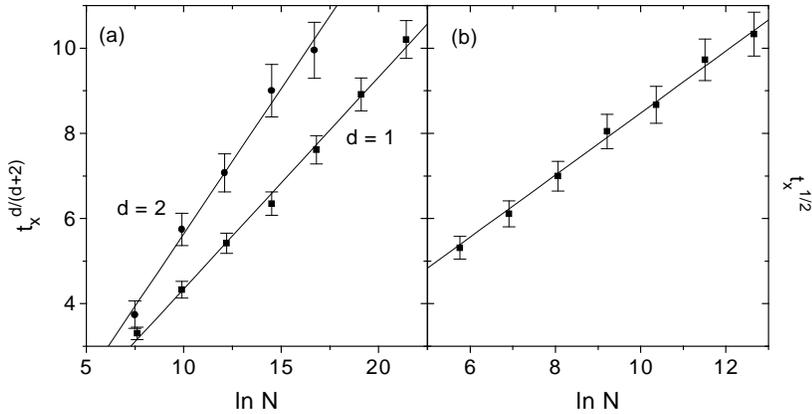,width=6cm,clip=,bbllx=140,bblly=200,bburx=440,bbury=750,angle=-90}
\caption{Plot of $t_\times^{\alpha/(\alpha+\gamma)}$
 as a function of $\ln N$,
for ({\bf a}) the trapping model and ({\bf b}) the hierarchical  model. 
  The straight
line supports (\ref{eq8}).
The crossover times $t_\times$ were obtained from
the positions of the minima of Fig.~\ref{fig02}}
\label{fig03}
\label{fig03a}
\label{fig03b}
\end{figure}

To show the dependence of the crossover time $t_\times$ on the system size $N$
we have plotted, in Fig.~\ref{fig03}, the values of $t_\times^{\alpha/(\alpha+\gamma)}$
 as a function of 
$\ln N$. The crossover time was obtained numerically from the 
 position of the minima of the 
curves in Fig.~\ref{fig02}. The resulting straight lines are in
full agreement with the prediction of (\ref{eq8}), supporting our
analytical approach.

In the following we discuss the relevance of our results to
Monte-Carlo simulations and experiments. There exists a
long standing puzzle
in Monte-Carlo simulations of the trapping problem in $d=2$ and 3, that
the predicted stretched exponential could not be observed \cite{bib08}, even
for survival probabilities $Q(t)/Q(0)$ down to
$10^{-21}$ in $d=2$ [8b] and $10^{-67}$ in $d=3$ [8g].

The finding of the logarithmic dependence of $Q(t)$ on the system
size $N$  explains this puzzle.
The Monte-Carlo simulations in $d=2$ and 3 were typically performed
 on $10^3$ configurations  with  about $10^4$ traps, which is
equivalent to having a single system with $N\sim 10^7$ traps.
  Using 
(\ref{eq10}), we expect for $N=10^7$ traps  $Q(t_\times)/Q(0)\cong 10^{-7}$ in
$d=2$.  Indeed, for times above $t_\times$ the exponent $\beta$ aproaches unity
as predicted by our theory and seen clearly in Fig.~\ref{fig02a}a. Moreover, for this
system size $\beta$ never reaches the predicted thermodynamic value $\beta =
0.5$, the minimum value of $\beta$ is about $0.65$.
For $d=3$, $Q(t_\times)/Q(0)\cong 10^{-11}$ thus for smaller survival values
($t>t_\times$) one again expects increasing values of $\beta$ approaching
unity.
This explains the exponential decay found
in the early  Monte-Carlo simulations. Our results show that this is not
an artefact but due to the  finite size of the system. 
Moreover, they clearly 
indicate that the  thermodynamic limit can not even be reached in
one-dimensional macroscopic systems.

It would be of interest to test the above
prediction experimentally by preparing experimental
realizations where size effects can be controlled.
Equations (8) and (10) suggest that the behavior
around the crossover can be measured experimentally..  For
the trapping problem in linear systems, which has been studied experimentally
\cite{bib27,bib28}, we expect for $10^8$ sites and concentrations of traps $c$
between
$10^{-4}$ and $10^{-2}$, that $Q(t_\times)/Q(0)\sim10^{-2}\div10^{-3}$,
which is a survival range that can be detected experimentally.
For the non identical particles (case (iii)), we expect for $10^8$ sites
and concentration of traps $c$ between $10^{-4}$ and $1$ that
$Q(t_x)/Q(0)\sim10^{-1}\div10^{-2}$ which is a survival range
that can be well detected experimentally.
The same arguments are valid for the target problem and therefore a similar
crossover from stretched exponential to exponential decay is expected in
relaxation experiments in low dimensional geometries \cite{bib29}.
Mesoscopic systems such as quantum dots, are also promising
candidates for experiments where the crossover can be relevant.
Identifying
the logarithmic size dependence in experiments
 may provide support to
the theories claiming that the  observed stretched
exponential is due to competing exponential processes, represented
by (\ref{eq2}).

This work was supported by the German Israeli Foundation (GIF).

%INDEX%%%%%%%%%%%%%%%%%%%%%%%%%%%%%%%%%%%%%%%%%%%%%%%%%%%%%%%%%%%%%%%
\clearpage
\addcontentsline{toc}{section}{Index}
\flushbottom
\printindex
%%%%%%%%%%%%%%%%%%%%%%%%%%%%%%%%%%%%%%%%%%%%%%%%%%%%%%%%%%%%%%%%%%%%%


\begin{thebibliography}{99}
\addcontentsline{toc}{section}{References}

\bibitem{bib01} R.~Kohlrausch, Ann Phys. (Leipzig) {\bf 12}, 393 (1847)

\bibitem{bib02} G.~Williams and D.~C.~Watts, Trans. Faraday Soc. {\bf 66},
 80 (1970)

\bibitem{bib03} V.~Chamberlin, G.~Mozurkewich and R.~Orbach, \PRL {\bf 52},
 867 (1984)

\bibitem{bib04} F.~Mezei and A.~P.~Murani, J. Magn. Mater. {\bf 14}, 211
 (1979)

\bibitem{bib05} A.~Plonka, The Dependent Reactivity of Species in Condensed
 Matter (Springer-Verlag, New York 1986)

\bibitem{bib06} A.~A.~Jones et al., Macromolecules {\bf 16}, 658 (1983)

\bibitem{bib07} K.~L.~Li et al., Macromolecules {\bf 21}, 2940 (1983)

\bibitem{bib08} (a) N.~D. Donsker and S.~R.~S. Varadhan, Commun. Pure Appl.
 Math. {\bf 32}, 721 (1979); (b) P. Grassberger and Procaccia, J.~Chem.
 Phys. {\bf 77}, 6281 (1982); (c) J.~Klafter, G.~Zumofen, A.~Blumen,
 J.~Phys. Lett. {\bf 45}, L49 (1984); (d) I. Webman, \PRL {\bf 52},
 220 (1984); (e) S.~Havlin, M.~Dishon, J.~E.~Kiefer, G.~H.~Weiss, \PRL
 {\bf 53}, 407 (1984); (f) M.~Fixman, \PRL {\bf 52}, 791 (1984);
(g) J.K. Anlauf,  \PRL {\bf 52}, 1845 (1984)


\bibitem{bib09} A.~K.~Jonscher,
Nature {\bf 267}, 673 (1977)

\bibitem{bib10} K.~L.~Ngai,
Comments Solid State Phys. {\bf 9}, 127 (1979); {\bf 9}, 141 (1980)

\bibitem{bib11} K.~Funke,
Prog. Solid St. Chem. {\bf 22}, 11 (1993)

\bibitem{bib12} J.~Klafter, M.~F.~Shlesinger,
Proc. Natl. Acad. Sci. U.S.A. {\bf 83}, 848 (1986)

\bibitem{bib13} H.~Scher, D.~Bendler and M.~Shlesinger,
Physics Today, January 1991

\bibitem{bib14} J.~C.~Phillips,
Rep. Prog. Phys. {\bf 59}, 1133 (1996)
   
\bibitem{bib15}  J.~C.~Rasaiah, J.~Zhu, J.~B.~Hubbard, and R.~J.~Rubin, J.
Chem. Phys. {\bf 93}, 5768 (1990);

\bibitem{bib16} R.~G.~Palmer, D.~L.~Stein, E.~Abrahams and P.~W.~Anderson,
\PRL {\bf 53}, 958 (1984)

\bibitem{bib17} W.~G\"otze and L.~Sj\"ogren,
Rep. Prog. Phys. {\bf 55}, 241 (1992)

\bibitem{bib18} M.~H.~Cohen and G.~S.~Grest,
Phys. Rev. B{\bf 24}, 4091 (1981)

\bibitem{bib19} M.~F.~Shlesinger and E.~W.~Montroll,
Proc. Natl. Acad. Sci.  U.S.A. {\bf 81}, 1280 (1984)

\bibitem{bib20} A.~Blumen, J.~Klafter, G.~Zumofen,
in Optical Spectroscopy of Glasses,
ed. I.~Zchokke (Reidel, Dordrecht 1986)

\bibitem{bib21} A.~Bunde, S.~Havlin, J.~Klafter, G.~Graff and A.~Shehter,
%%%%Anomalous size dependence of relaxational processes,
\PRL {\bf 78}, 3338 (1997)

\bibitem{bib22} J.~C.~Phillips, J.~C.~Rasaiah and J.~B.~Hubbard,
%%%%Comment on "Anomalous size dependence of relaxational processes",
\PRL, {\bf 80} 5453 (1998);
A.~Bunde, S.~Havlin and J.~Klafter,
%%%%Comment on "Anomalous size dependence of relaxational processes" - Reply,
\PRL, {\bf 80} 5454 (1998)

\bibitem{bib23} A.~Bunde, L.~L.~Mosely, H.~E.~Stanley, D.~Ben-Avraham
and S.~Havlin,
{\it Phys. Rev.\/} A{\bf 34}, 2575 (1986)

\bibitem{bib24} F.~Sciortino and P.~Tartaglia,
%%%%Cluster aggregation under diffusion,
Physica A {\bf 231}, 191 (1996)

\bibitem{bib25} T.~E.~Harris,
J. Appl. Pro. {\bf 2}, 323 (1965);
P.~M.~Richards,
Phys. Rev. B{\bf 16}, 1393 (1977);
S.~Alexander and P.~Pincus,
Phys. Rev. B{\bf 18}, 2011 (1978)

\bibitem{bib26} A.~Bunde nad J.~Dr\"ager,
Phys. Rev. E{\bf 52}, 52 (1995);
J.~Dr\"ager and A.~Bunde,
{\bf ibid}, {\bf 54}, 4596 (1996)

\bibitem{bib27} R.~A.~Auerbach and G.~L.~McPherson,
Phys. Rev. B{\bf 33}, 6815 (1986)

\bibitem{bib28} R.~Knockenmuss and H.~U.~Gudel,
J. Chem. Phys. {\bf 86}, 1104 (1987)

\bibitem{bib29} J.~M.~Drake et al.,
\PRL {\bf 61}, 865 (1988)

\end{thebibliography}
\end{document}